# Discovering Functional Zones Using Bus Smart Card Data and Points of Interest in Beijing

Haoying Han, Zhejiang University

Xiang Yu, Zhejiang University

Ying Long, Beijing Institute of City Planning

**Abstract** Cities comprise various functional zones, including residential, educational, commercial zones, etc. It is important for urban planners to identify different functional zones and understand their spatial structure within the city in order to make better urban plans. In this research, we used 77976010 bus smart card records of Beijing City in one week in April 2008 and converted them into two-dimensional time series data of each bus platform, Then, through data mining in the big database system and previous studies on citizens' trip behavior, we established the DZoF (discovering zones of different functions) model based on SCD (smart card Data) and POIs (points of interest), and pooled the results at the TAZ (traffic analysis zone) level. The results suggested that DzoF model and cluster analysis based on dimensionality reduction and EM (expectation-maximization) algorithm can identify functional zones that well match the actual land uses in Beijing. The methodology in the present research can help urban planners and the public understand the complex urban spatial structure and contribute to the academia of urban geography and urban planning.

## 1 Introduction

Ae city, whether from a material or a social level, is a complex space system (Lai and Han 2009; Batty 2009), and the study based on the microscopic objects is an important way to understand the operation of this complex system. Due to the limitation of data, the classic study of urban elements, organization and structure often limited in large scale (such as townships or traffic analysis zones). But with the advent of the era of big data, Ubiquitous personal volunteered data have provided a new channel to describe and understand the structure of urban space (Batty 1990).

The big data, which has been a big concern in all research areas, mainly consists of bus smart card records, flight records, bank card records, twitter records, mobile phone records and so on, which are provided by Location Based Services(LBS) like Global System for Mobile(GSM), Global Positioning System(GPS), Social Networking Services(SNS) and Wi-Fi. These data can be formed into a log of citizens' trip behavior (Neuhaus 2009), and can be used to real-time monitor the urban activities (Calabrese and Ratti 2006), analyze the intensity and temporal distribution of urban activities (Ratti et al.2006) and study human mobility pattern in urban(Gonzalez et al.2008).

Human Mobility has a close relationship with urban spatial structure (Goodchild and Janelle 1984;

Goodchild et al.1993). In existing researches, urban spatial structure is often used by researchers to study human activity, analyze citizens' trip behavior and explore the impact of urban spatial structure on people' traveling(Jiang et al.2009). For example, by studying the structure of urban land-use, urban commuter model was studied (Hamilton 1982), and the impact of the spatial structure on residents' commuting behavior was analyzed (Liu and Wang 2011; Wang and Chai 2009). However, few studies have discussed how to use the existing data of urban human activity for relevant studies of urban spatial structure which can be very important. Because along with the development of the city, urban land use and spatial structure are changing rapidly and the city is developing from the previous single-center model to a multi-center model (Anas et al.1998; McMillen and McDonald 1997). The immediate and clear division of urban functional areas can give inspiration to city planners on the future planning, and validate the land use planning of the past. However, traditional studies on land use and urban spatial structure are mainly based on remote sensing data(Lu and Weng 2005; Xiao et al.2006) which is expensive and lacks timely updates, thus cannot meet the needs of urban planners and scholars. Therefore, it will become a top research to use the vast amounts of data provided by LBS to analyze human activity and carry out research of urban spatial structure.

In existing researches of urban spatial structure, GSM and GPS data is most widely used. For example, Qi et al. used GPS information of driving taxis in Hangzhou to analyze the relationship between passengers' patterns of getting on and off and the social function of urban areas (Qi et al.2011). Based on one week's GPS data from more than 6600 driving taxis in Shanghai, Liu et al. used the "source-sink" model proposed by Pulliam to characterize the daily traffic model, and then to analyze the present situation of land use in Shanghai(Liu et al.2012; Pulliam 1988). Yuan used GPS information of taxi and urban POI (Point of Interests, POIs) data to establish a semantic model and study the functional partition of different regions of the city with the help of data mining method(Yuan et al.2012).

In recent years, SCD (Smart Card Data) has gradually been used in the study of urban as a kind of large-scale data with spatiotemporal labels. Sun analyzed passengers' spatio-temporal density and activity tracks based on the SCD data of Singapore (Sun et al.2012). Joh and Hwang used 10 million pieces of SCD data of Seoul metropolitan area to analyze the feature of cardholders' routes and urban land use (CH and Hwang 2010). Long et al. analyzed the relationship between working and living locations as well as the feature of commuting direction using SCD data of Beijing (Long et al.2012). In addition, some researchers have begun to use POIs (Point of Interests) data in the study of urban space. POIs are some basic locations of the city and mainly include buildings with landmark function in the local area. Since to a large extent, POIs can enhance the ability to describe the physical location and improve the accuracy and speed of the geographical location, it has been widely applied to the study of urban spatial structure. According to the significant degree of difference, Zhao et al. extracted hierarchical landmarks from POI data, and obtained the hierarchical knowledge space that can be used in intelligent route guidance (Zhao et al.2011). Since LBS technology is featured with high positioning accuracy, interactive feature and huge amount of data while POI has obvious advantages in identifying geographical pattern, its id of great significance to integrate the two in the study of urban spatial structure.

Information obtained by LBS technology is still in its original state, data. Because the amount of data is too large, it is usually hard to extract useful information using the traditional data analysis method and technology (Witten and Frank 1999). In recent years, with the rapid development of computer techniques, database management system and artificial intelligence technology gradually became mature and their combination has contributed to produce a new technology, data mining, which realizes the effective characterization and analysis of large data(Tan et al 2006). Scholars also have begun to try to use the classification, association analysis and cluster analysis in data mining technique to discovery potentially and useful information from vast amounts of GIS data. For the past few years, cluster analysis has been widely used at home and abroad, to analyze daily activities of urban people like working, living, attending school, travelling and shopping based on GPS, GSM, SCD and other data (Jiang et al.2012; Sun et al.2011), and thereby to identify the space-time structure of the city (Jiang et al 2012) as well as the immediate and detailed information of urban land use (Pan et al.2013).

In actual studies, the data provided by LBS technology is often continuous in time and such data are called time-series data (Agrawal et al.1993). Time-series data are usually massive data that may get a lot of noise in it and have poor efficiency or even impossibility in direct cluster analysis of raw data. Therefore, dimensionality reduction and feature transformation for multi-dimensional time series data is needed. Discrete Fourier Transform (DFT) (Agrawal et al.1993), Principal Component Analysis (PCA)(Sun et al.1994), Singular Value Decomposition (SVD) (Korn et al.1997) are some commonly-used methods for this. Cluster analysis of time series data can be similarity-based, feature-based, model-based and segmentation-based. And the choice mainly depends on the type of application data and the purpose of cluster analysis (Jiang et al.2005). Traditional cluster analysis is mostly vector-based which cannot well solve the problem of time series clustering. In recent years, model-based clustering was more used in the study on clustering analysis of time series.

This study is to establish the model of Discovering Zones of different Functions (DZoF) based on the original data of SCD and POIs. In this model, the bus platform traffic data model was developed. We have used the model-based algorithm —expectation maximization(EM) in cluster analysis of 8691 bus platforms in Beijing, as well as built the pattern recognition rules of SCD data mining based on the traditional studies of residents' commuting behavior, general cognition and POIs model, and conducted a functional interpretation of clusters resulted from previous clustering analysis. According to DZoF model, this study ultimately determined the function of each bus platform in Beijing, and made a summary on a scale of TAZ to achieve recognition of the function of different regions. In order to verify the validity of DZoF(Discovering Zones of different Functions) model in identifying the result, the chapter also made a contrastive analysis between the land-use status map of the overall urban planning of Beijing(2004-2020) and Google map of the area.

The methods section of this chapter gives a definition of DZoF model used for urban functional identification, and introduces the specific method to use DZoF model in the cluster analysis of multi-dimensional time series, and ultimately identify the practical meaning of clustering result. In the application section, Beijing is taken for an example, a continuous week's bus smart card records of Beijing in April 2008 and urban POI data are used to identify each functional area of the city while experimental result is also tested. The final part is about the summary and discussion of the entire study.

## 2 Overview of study area and explanation of data

### *2.1 Overview of study area*

The study area of this paper is Beijing city area, with a total area of 16,410 square kilometers and a permanent population of 20,693,000[1]. Beijing has a modern and tridimensional transportation network which ramifies all over the city. Till 2008, there are 17,000 km bus-lines and 20600 operating vehicles, 8 lines of rail transit with 200 km of operating mileage, and 66,000 operating taxis[2].

### *2.2 Data*

#### 2.2.1 Lines and platforms of buses in Beijing

The main data of this study mainly is a continuous week (April 7th-April 13th)'s bus smart card records of Beijing in 2008 (not including smart card records of rail transit), which covers more than 600 bus lines (a total of 1,287 inbound and outbound records which contains 566 records of one-ticket lines and 721 of segmented-pricing lines), and about 37,000 bus stops[3] and 8,691 bus platforms. Figure 1 shows the distribution of bus platforms in Beijing.

---

[1] The data is from the 2012 Statistical Yearbook of Beijing(http://www.bjstats.gov.cn).

[2] The data is from the statistics on Beijing Transportation website(http://www.bjbus.com/)

[3] It is not the number of but platforms, but the sum of bus stops of all bus lines. A bus stop is the name of a platform in a bus line.

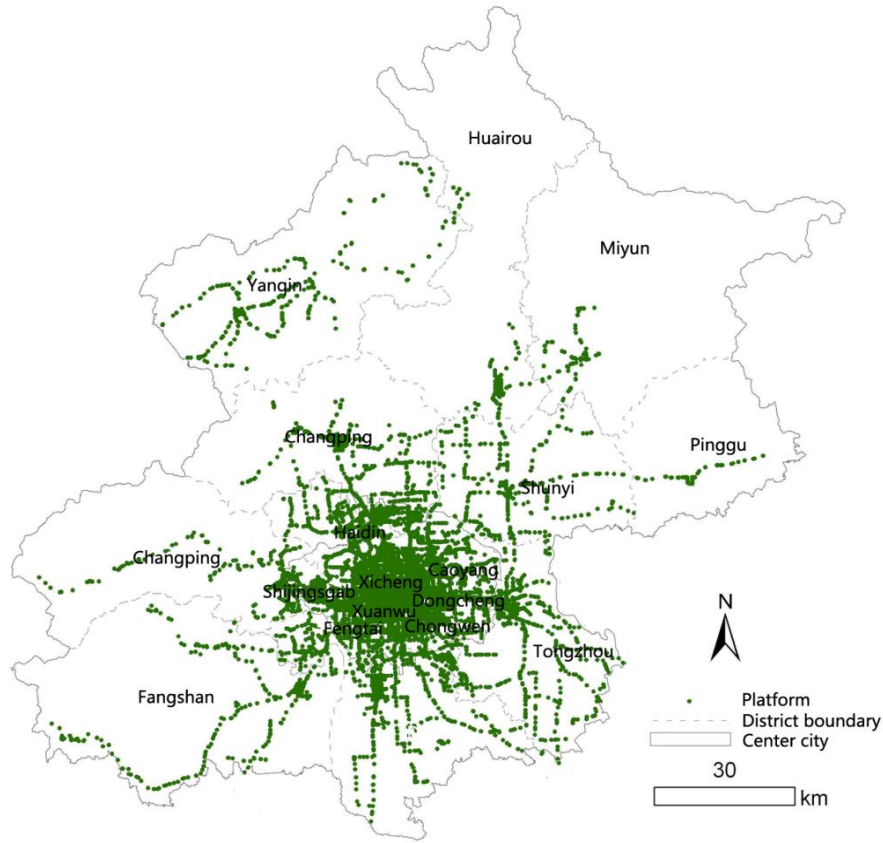

**Fig.1** The bus platforms of the Beijing Metropolitan Area (BMA)

**2.2.2 Smart Card Data**

Due to the non-technical factors, the smart card data used in this study does not include those of bus and rail transit lines of Xianglong Company. The basic information that each record contains includes: card-swiping time and location of each card holder (the location is represented by the number of the line and the stop), type of card (ordinary, student or staff card), and transaction sequence number (representing the cumulative number of a cardholder's swiping card), driver number and vehicle number. There are 77,976,010 card-swiping records in the week.

Beijing bus lines have two ways of pricing: (1)one-ticket lines with short distance mostly located in the city centre and SCD of this kind only records passengers' pick-up time rather than drop-off time; (2)segmented-pricing lines, usually with long route and both terminal and starting station locating outside the Fifth Beltway. SCD of this kind records the complete spatial-temporal information of cardholders' behavior of card swiping. In order to collect both pick-up and drop-off flow, most of the data used in this study is of segmented-pricing lines, which has 37,649,207 records in total.

**2.2.3 Point of interest**

The POI data used in this study is collected from Beijing in 2010 and has a total of 113,810 records, obtaining from Sina Micro-blog Geographic Service Platform [4](as shown in figure 2).

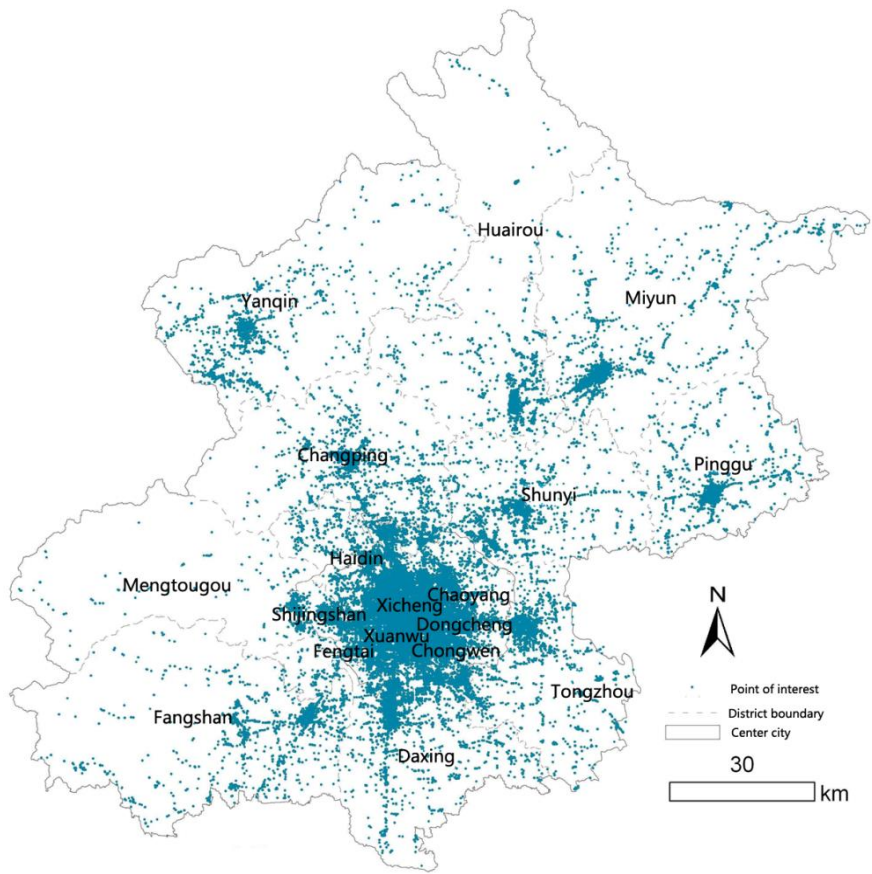

**Fig. 2** the POIs of the Beijing Metropolitan Area (BMA)

The classification and explanation of POIs is shown in table 1.

**Table 1** Codes, Categories and Description of POIs

| 1st level code | Classification of POI | Explanation | 1st level code | Classification of POI | Explanation |
|---|---|---|---|---|---|
| 01 | Automobile service | Oil station, gas station, automobile care, car-washing station, car renting, etc. | 11 | Place of interest | Park plaza, scenic spot |
| 02 | Vehical sales | VW, Toyota, Honda, GM and BMW sales. | 12 | Business and residential area | Industrial park, , residential area, etc. |
| 03 | automobile maintenance | automobile integrated maintenance, maintenance of VM, Honda, etc. | 13 | Government agency and social organization | Government agency, foreign institution, social organization, etc. |

---

[4] Sina Micro-blog LBS Platform, officially opened in 2012 April, provides third party developers with free access to Sina location service. its most outstanding part are the two functions that base on user and POI. Related interface based on user can allow users to obtain individual's dynamic time line. and POI interface is based on a specific location.(Http://open.weibo.com/)

| 04 | Motorcycle service | Motorcycle sales and maintenance | 14 | Science, educational and cultural service | Museum, library, cultural center, school, research institution, etc. |
| 05 | Catering service | Chinese and foreign restaurants, fast-food restaurant, coffee house, etc. | 15 | Transport facility | Airport, train station, harbor and wharf, subway station, etc. |
| 06 | Shopping service | Shopping mall, convenience shop, household appliance store, supermarket, home furnishing store, etc. | 16 | Finance and insurance service | Bank, insurance company, securities company, finance corporation. |
| 07 | Life service | Travel agency, post office, logistic express, talent market, electric power office, beauty salon, etc. | 17 | Corporation | Well-known enterprise, company, factory, agricultural base, etc. |
| 08 | Sport and leisure service | Stadium, entertainment venue, leisure facility, movie theater, etc. | 18 | Road affiliated facility | Toll station, service center of gas station. |
| 09 | Health care service | General and specific hospital, clinic, etc. | 19 | Address information | Transport-related place name, city center. |
| 10 | Accommodation service | Hotel, guest house. | 20 | Public facility | Newsstand, public toilet, shelter. |

The amount of each category of POI is shown in Figure 3.

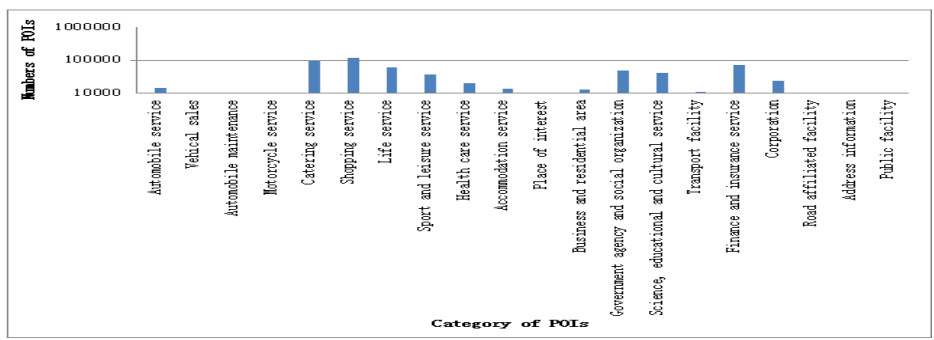

**Fig. 3** POIs counts for each category in 2010

## 3 Methods

In this study, the identification of city functional regions was achieved mainly through the construction of DZoF (Discovery Zones of different Functions) model.

First, by using SQL Server, acquire and pre-process SCD and POIs data, Platform Flow model (PF model) and POIs data model were built. And by using the cluster analysis techniques of multi-dimensional data, we designed the feature of one week's traffic data, use the EM algorithm for clustering analysis and get clusters. And again, the function of the result clusters was interpreted in three aspects: POI model, feature of residents' commuting behavior and residents' general cognition; the function of each bus platform was defined. Public management, science, educational and cultural service, residential, commercial and entertainment function, and scenic spot are some main functions. The process flow of this study is as shown in figure 4.

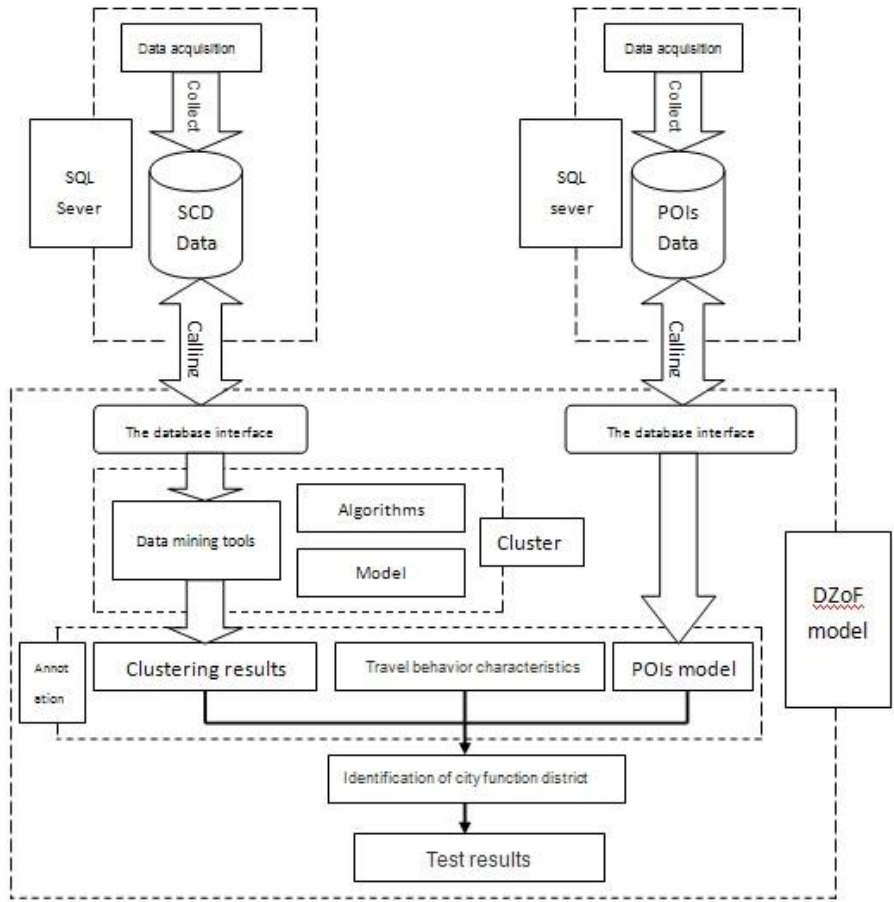

**Fig. 4** The process flow diagram of this chapter

## 3.1 Blind clustering of bus platforms

### 3.1.1 Pre-processing of SCD

Flow statistics collected according to bus lines compose the raw data of the study, which has a problem of having many bus lines going through the same platform. By using SQL Server, statistics of different bus lines which go through the same platform were added together, resulting in the flow statistic of each bus platform, $f_{x,y,z}$ (x is Platform ID, x=1,2,…,8691; y represents date, y=7,8,…,13; z represents time, z=0,1…,23).

### 3.1.2 PF (Platform flows) data model

For each bus platform, an inflows vector was built ($X_{7,0}$, $X_{7,1}$, …, $X_{i,j}$, …, $X_{13,23}$). $X_{i,j}$ represents the number of passengers that have been picked up on the platform in the $j^{th}$ hour on April $i^{th}$ 2008 (i=7, 8,…,13; j=0, 1…,23). While at the same time, a outflows vector was built ($Y_{7,0}$, $Y_{7,1}$,…,$Y_{i,j}$,…,$Y_{13,23}$). $Y_{i,j}$ represents the number of passengers that have been dropped off on the platform in the $j^{th}$ hour on April $i^{th}$

2008( i=7,8,…,13, j=0,1…,23).

By transforming the data into two-dimensional time-series data, making dimensionality reduction and constricting a linear function, the study uses a ratio of the number of pick-up passengers to the number of drop-off ones at different time as an index for comparing the similarity of platform flows:

$$Z_{ij} = \frac{X_{i,j}}{Y_{i,j}} \qquad . \qquad (1)$$

($X_{i,j}$ represents the number of passengers that have been picked up on the platform in the $j^{th}$ hour on April $i^{th}$; $Y_{i,j}$ represents the number of passengers that have been dropped off on the platform in the $j^{th}$ hour on April $i^{th}$)

Thus, for each platform, a PF model was built ($Z_{7,0}$, $Z_{7,1}$,…, $Z_{i,j}$,…,$Z_{13,23}$). $Z_{i,j}$ represents the ratio of passengers that have been picked up on the platform in the $j^{th}$ hour on April $i^{th}$.

### 3.1.3 Data dimensionality reduction

From the statistics of average inflows and outflows in various departure hours of a week (as shown in Figure 5), it was observed that people's trip hour converges on 5:00-23:00 (above the red line in Figure 5). After removing redundant features of the raw data, PF data decreased from 168d to 126d.

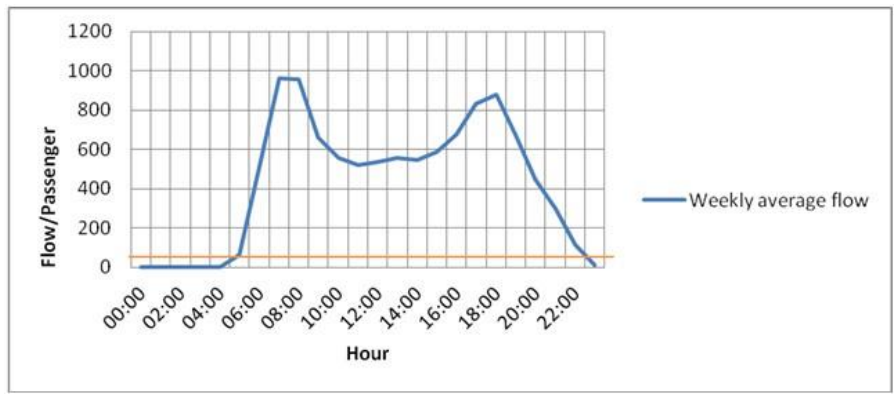

**Fig. 5** The average passenger flows in various departure hours of a week

Next, we drew a graphics of the inflows in various departure hours of one week of the No.1934 platform (Figures 6, 7). From the graphics, it is observed that statistics of weekdays have strong consistency while statistics of two days in weekends have strong consistency.

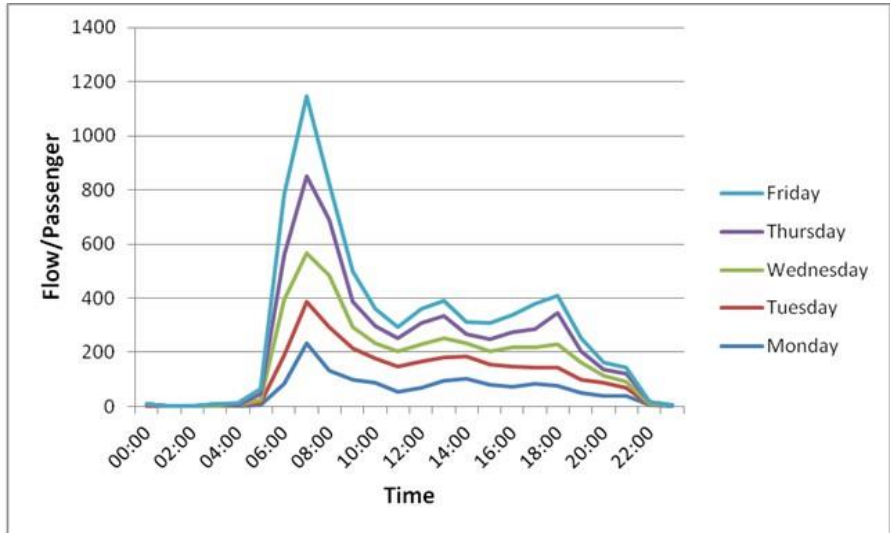

**Fig. 6** The inflows in various departure hours on weekdays of No.1934 platform

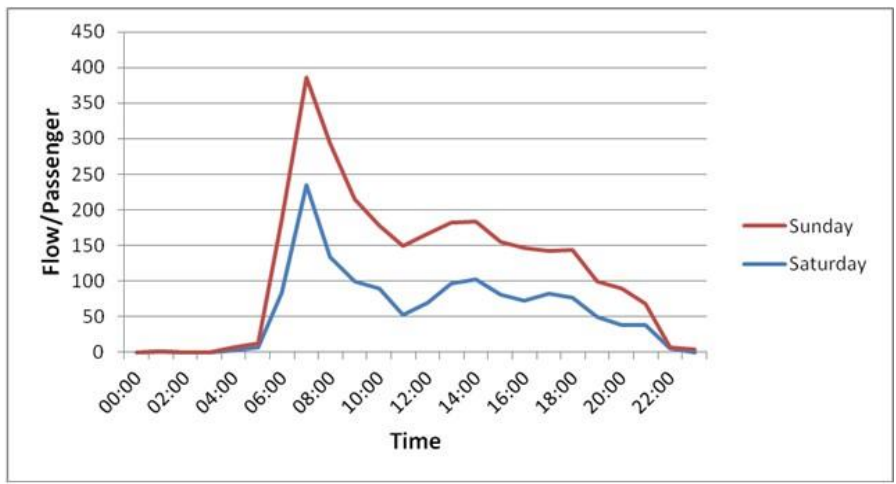

**Fig. 7** The inflows in various departure hours on the weekend of No.1934 platform

Then, correlation analysis of statistics of platform flows on weekdays and weekends was made and Pearson correlation coefficient between each two attributes was calculated. The results show that the flow data in the same hour of different weekdays was significantly correlated at 0.01 (bilateral) levels Therefore，we designed the feature of flow data and calculated the arithmetic mean of statistics of weekdays and weekends.

### 3.1.4 Expectation-Maximization (EM) algorithm

In this paper, expectation maximization (EM) algorithm was chosen as the method for clustering on bus platforms. In this algorithm, for each object, its probability of each distribution is calculated, which is equivalent to a procedure in K-means algorithm that assigns each object to a cluster; and maximum likelihood estimating in EM algorithm is equivalent to cluster's centered calculating in K-means algorithm. But compared with the K-means algorithm, EM algorithm is more general and can be applied to different classes and find clusters of different sizes. At the same time, EM algorithm is model-based, which can eliminate the complexity of data association.

EM algorithm is a clustering method that uses hybrid model. The principle of model-based clustering is to assume that data is obtained by a statistical process which can be classified with a statistical model. Therefore, a statistical model that best fits the data can be found and parameters of the model can be estimated from data. The basic process of EM algorithm can be summarized as: first, do the initial guess of parameters; then, iteratively refine the estimation. Estimation of parameters in the algorithm is carried out by using the maximum likelihood method. Probability density of points generated from one-dimensional Gaussian distribution is:

$$\text{prob}(\chi|\Theta) = \Pi \frac{1}{\sigma\sqrt{2\pi}} e^{-\frac{(x_j-\mu)^2}{2\sigma^2}} \qquad . \tag{2}$$

If the value of σ and μ is unknown, then it requires a process to estimate them, which means choosing σ and μ that can maximize the formula above. This way of estimating model parameters is called maximum likelihood estimation in statistics.

## *3.2 Identification of urban functional areas*

### 3.2.1 Collection of POI data of bus service area

In this study, service area of bus platform is defined the area within a radius of 500 meters (Huang 2006) around the bus platform. For each bus platform, the number of POI data of different categories is calculated, indicated as . i is platform ID(i=1,2,…,8691); j is 1st level code of POI (j=1,2,…,20).

### 3.2.2 Data standardization

Among POIs data of Beijing in 2010, there are 90,819 records of catering service POI, 116,499 shopping service POIs and only 4,575 POIs of place of interest. In the process of statistical analysis, function identification will be affected by difference in POIs's magnitude.

Therefore, original POIs data need to through Z-Score standardization according to the following formula:

$$x_{ij}^* = \begin{cases} \frac{x_{i,j}-\bar{x}_j}{S_j} & (S_j \neq 0) \\ 0 & (S_j = 0) \end{cases} \tag{3}$$
$$(i = 1,2,\dots,n; j = 1,2,\dots,m) \quad .$$

### 3.2.3 POI data model

For each platform, a POIs feature vector, FD (Frequency Density) model was built, denoted as $f_{d1}$, $f_{d2}$, …, $f_{d20}$.

$f_{di}$ denotes the frequency density of the $i^{st}$ category of POI in platform service area R:

$$\text{fd}_i = \frac{\text{The standardized number of the ist category of POI of the platform}}{\text{The area of platform service arear r}} \quad (4)$$

Similarly, for each platform, another POI feature vector, CR (Category Ratio) model was built, denoted as $cr_1, cr_2,\ldots, cr_{20}$.

$cr_i$ denotes the percentage of the $i^{st}$ category of POI in all POI of the area:

$$\text{cr}_i = \frac{\text{The standardized number of the ist category of POI of the platform}}{|\text{The number of POI in platform service arear r}|} \quad (5)$$

**3.2.4 Urban functional identification**

The study used traditional rules of data acquisition, and then used these rules for pattern recognition of large data, which was to recognize the function of the clusters obtained from blind clustering of bus platforms by using the correlation between residents' trip time and purpose, residents' general cognition and POI model.

Function recognition is a comprehensive application of the following three methods:

(1) Calculate each cluster's FD (Frequency Density) model, and sort the results (achieving internal ranking); second, calculate each cluster's CR (Category Ratio) model and sort the results (achieving external ranking).

(2) Identify the feature of flows in trip hours of each cluster.

(3) Residents' general cognition. People usually are aware of the function of some well-known places, such as the Imperial Palace, Chinese Silicon Valley and Beihai Park.

## 4 Results

### *4.1 Clustering results of bus platforms and Summary at TAZ scale*

After clustering of bus stops based on flow data by using EM algorithm in the software Weka, 6 different clusters were obtained (each bus stop only belongs to one cluster, C0-C5). Next, according to the spatial subordination between platforms and traffic analysis zones (TAZ), statistical work was carried out for each traffic analysis zone, and the cluster that exists most in a TAZ represents the category of the TAZ. The clustering results were summarized at TAZs scale (sparse for unclassified area), as shown in figure 8.

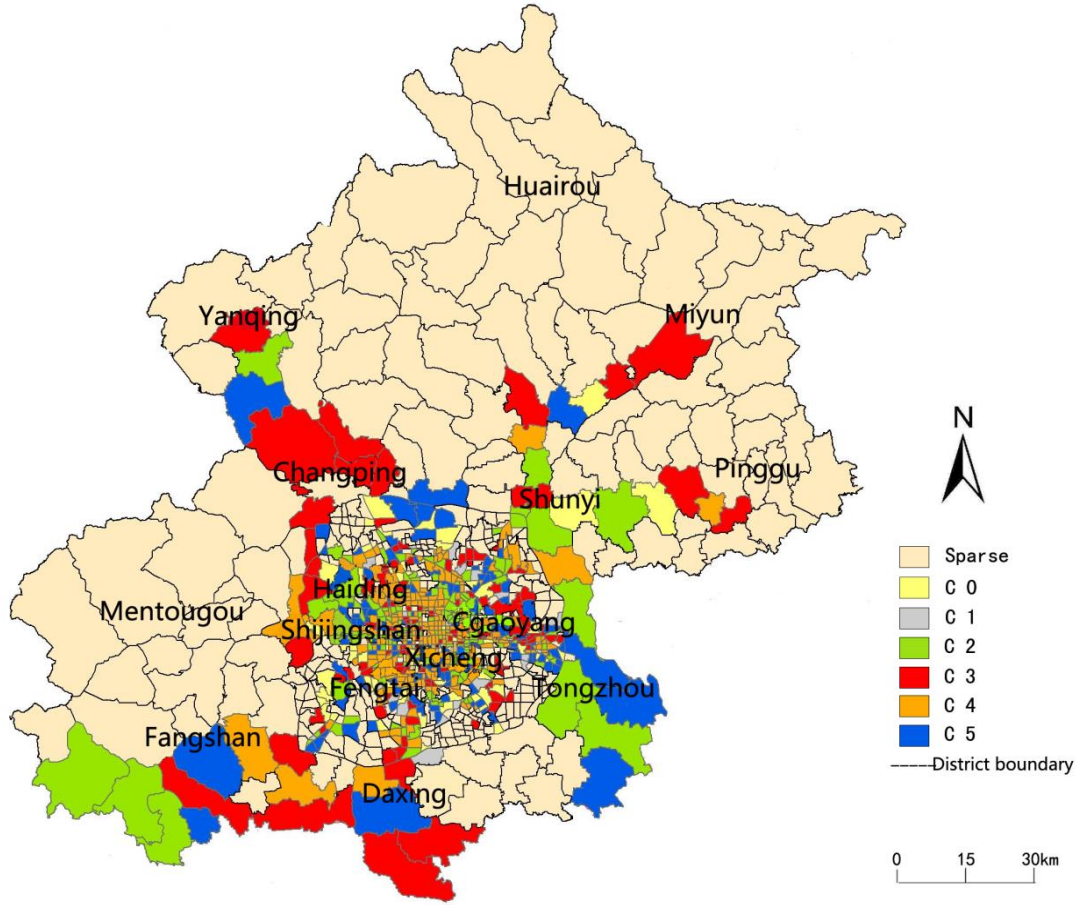

**Fig.8** Functional regions of the Beijing

## 4.2 Function Identification

### 4.2.1 Construction of POIs model

According to the results of blind clustering of platforms in 4.1, a POIs model was built for each cluster (C0-C5), and the value of FD (Frequency Density) and RCR (Rank of Category) of each functional region were calculated, as shown in Table 2.

**Table 2** Overall POI feature vector and ranking of functional regions by DZoF (the color of a cell indicates its value)

|  | C0 | | C1 | | C2 | | C3 | | C4 | | C5 | |
|---|---|---|---|---|---|---|---|---|---|---|---|---|
| POI | FD | RCR | FD | RCR | FD | RCR | FD | RCR | FD | RCR | FD | RCR |
| Automobile service | -0.077 | 7 | -0.025 | 6 | 0.073 | 9 | 0.03 | 19 | 0.021 | 18 | -0.02 | 6 |
| Vehicle sales | -0.075 | 6 | 0.034 | 2 | -0.006 | 19 | 0.089 | 14 | 0.073 | 13 | -0.063 | 12 |
| Automobile maintenance | -0.005 | 3 | 0.032 | 3 | 0 | 18 | 0.119 | 9 | 0.084 | 12 | -0.012 | 4 |
| Motorcycle service | 0.063 | 1 | 0.006 | 5 | 0.057 | 13 | 0.085 | 15 | 0.041 | 16 | 0.117 | 1 |
| Catering service | -0.186 | 18 | -0.109 | 13 | 0.142 | 1 | 0.149 | 7 | 0.205 | 5 | -0.095 | 18 |
| Shopping service | -0.173 | 16 | -0.141 | 16 | 0.039 | 15 | 0.214 | 3 | 0.107 | 11 | -0.051 | 10 |
| Life service | -0.156 | 13 | -0.157 | 18 | 0.099 | 5 | 0.216 | 2 | 0.114 | 10 | -0.026 | 7 |

| Sport and leisure service | -0.16 | 14 | -0.114 | 14 | 0.124 | 2 | 0.095 | 10 | 0.307 | 1 | -0.057 | 11 |
|---|---|---|---|---|---|---|---|---|---|---|---|---|
| Health care service | -0.106 | 9 | 0.013 | 4 | 0.06 | 12 | 0.187 | 5 | 0.056 | 15 | -0.004 | 3 |
| Accommodation service | -0.183 | 17 | -0.14 | 15 | 0.075 | 8 | 0.187 | 4 | 0.18 | 6 | -0.034 | 8 |
| Place of interest | -0.129 | 12 | -0.076 | 11 | 0.042 | 14 | -0.033 | 20 | 0.167 | 8 | -0.075 | 15 |
| Business and residential area | -0.073 | 5 | -0.094 | 12 | 0.072 | 10 | 0.152 | 6 | 0.07 | 14 | -0.018 | 5 |
| Government agency and social organization | -0.124 | 11 | -0.18 | 20 | 0.082 | 6 | 0.135 | 8 | 0.224 | 2 | -0.11 | 20 |
| Science, educational and cultural service | -0.202 | 19 | -0.173 | 19 | 0.068 | 11 | 0.067 | 18 | 0.22 | 3 | -0.095 | 17 |
| Transport facility | -0.173 | 15 | -0.076 | 10 | 0.111 | 4 | 0.089 | 13 | 0.17 | 7 | -0.066 | 13 |
| Finance and insurance service | -0.214 | 20 | -0.057 | 8 | 0.114 | 3 | 0.094 | 11 | 0.216 | 4 | -0.105 | 19 |
| Corporation | -0.12 | 10 | -0.144 | 17 | 0.075 | 7 | 0.069 | 17 | 0.117 | 9 | 0.017 | 2 |
| Road affiliated facility | -0.008 | 4 | -0.039 | 7 | 0.031 | 16 | 0.092 | 12 | -0.036 | 20 | -0.068 | 14 |
| Address information | 0.015 | 2 | 0.039 | 1 | 0.016 | 17 | 0.084 | 16 | 0.021 | 17 | -0.079 | 16 |
| Public facility | -0.102 | 8 | -0.068 | 9 | -0.048 | 20 | 0.214 | 1 | -0.036 | 19 | -0.044 | 9 |

### 4.2.2 The feature of residents' traffic flows

The features of flows (pick-up and drop-off number) on weekdays and weekends in a week of clusters clustered by EM are shown as Figure 9-12.

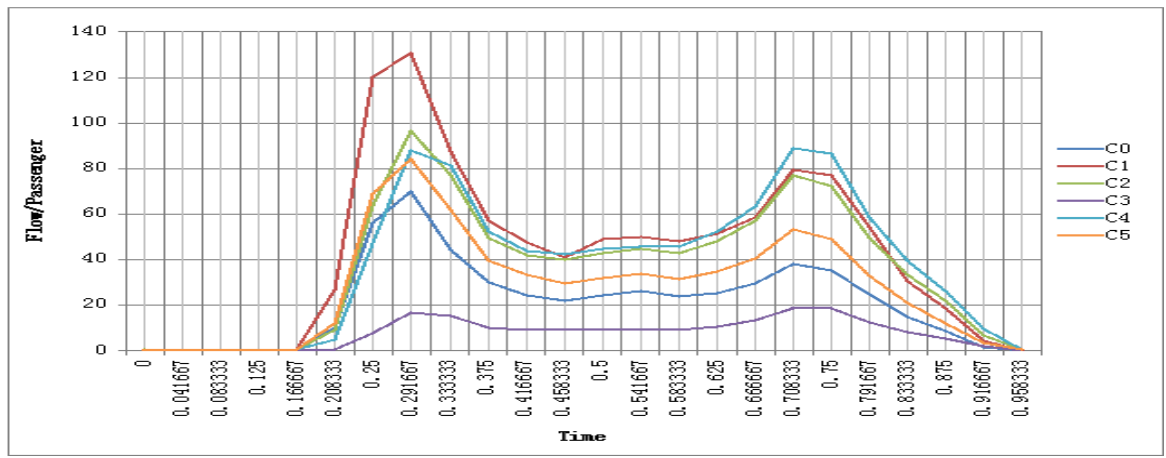

**Fig.9** The inflows on weekdays of clusters clustered by EM

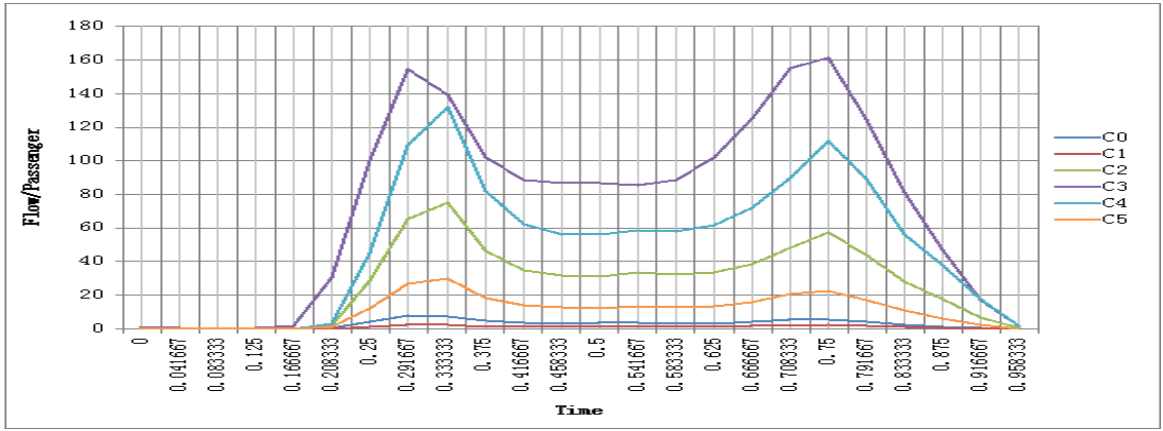

**Fig. 10** The outflows on weekdays of clusters clustered by EM

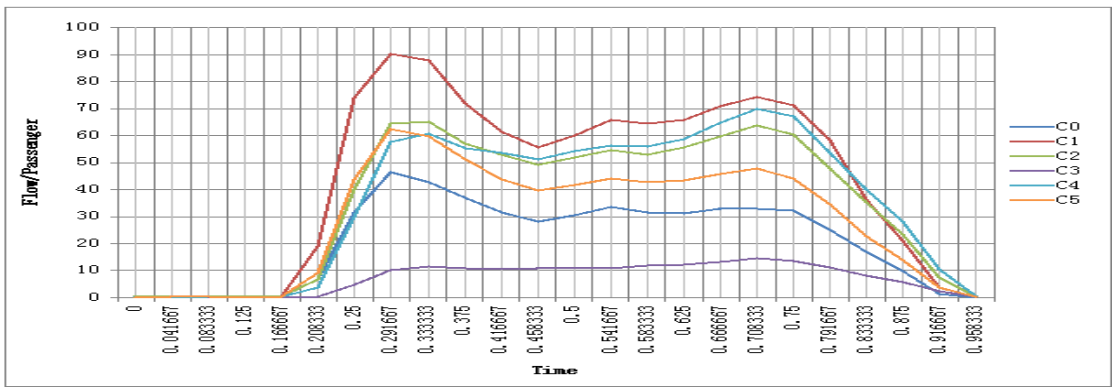

**Fig. 11** The inflows on weekends of clusters clustered by EM

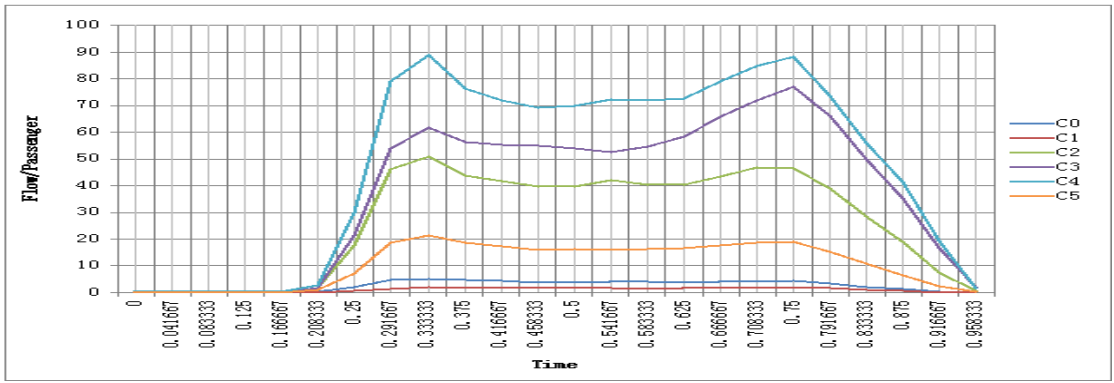

**Fig. 12** The outflows on weekends of clusters clustered by EM

### 4.2.3 Identification of results

Functions of clustering results of EM algorithm were identified as followed:

  a. Mature residential area (C0)

  Regions of this kind have got widely distributed residential POI and a high proportion of serviced apartment(FD=-0.073), while services for residents' life like health care, accommodation and education service is distributed in equilibrium, which have shown the feature of POI distribution of a typical residential area.

  At the same time, through the analysis of flow data of the week, it was observed that flows of departure in weekdays of the region peaked in the early morning(7-8 o'clock in morning rush hours) and the peak of flows of return appeared in the evening (17-19 o'clock in afternoon rush hours), which have shown the trip mode of a typical residential area.

 b. developing area(C1)

POIs of regions of this kind are mostly of motorcycle and car service category where 4S shops, motorcycle sales, automobile and motorcycle maintenance stations are relatively widespread and the surrounding infrastructure need to be improved.

 c. Scenic spots (C2)

Places of interest account for the highest proportion of POI in these regions. Compared with other kinds, regions of this kind have higher FD(Frequency Density) value(0.042) and are high-ranked in the external ranking of catering and accommodation service for tourists. And there is little difference between traffic flows of weekdays and weekends in these regions. Flows were relatively averaged in different hours of a day and there were usually higher flows in weekends than weekdays.

 d. Commercial and entertainment area (C3)

Regions of this kind have high FD value in catering service, shopping service and life service, respectively ranking second, first and first in all clusters. Meanwhile, the regions contain a high proportion of POI of catering service (with higher CR value. For example, clusters of catering service rank the 7th in CR value while clusters of shopping service rank the third). And form the above figures of traffic flows, it was observed that the peak of drop-off flows appeared in afternoon rush hours (17:00-19:00) on weekdays in these regions, which have shown that many people shopped and took part in recreational activities in these areas.

 e. Area of public management, science, education and culture (C4)

Regions of this kind have the highest percentage of POI of government agency and social association which have higher FD value (0.22) compared with other regions. POI of this kind account for 9.7% of POI in these regions, ranking second in RCR value. And there are many POI of science, educational and cultural service in these regions where transport facility, sport and leisure service and accommodation service are high-ranked in the external ranking.

 f. New residential area (C5)

Regions of this kind have similar POI data structure with C0. While ranking according to the percentage that each kind of POI account for, residential service ranks the 5th, health care service ranks the third and life service ranks the 7th in these regions, which have shown the feature of POI distribution of a typical residential area.

On the other hand, form the data of traffic flows, it was observed that flows of departure in weekdays of the regions peaked in the early morning(7-8 o'clock in morning rush hours) and the peak of flows of return appeared in the evening (17-19 o'clock in afternoon rush hours). But compared with C0, these regions had smaller flows which were about 1/4 of flows of C0, which have shown that these regions don't have big traffic flows and still in a stage of developing.

 g. Unclassified area (sparse)

Some regions lack data of traffic flow for being covered with mountains, forests, rivers and others, so there are put in one category in this study.

According to the results of functional identification, the area and population of each functional area was collected, as shown in Table 3.

**Table 3** Information of each cluster

| Functional area | Number of TAZ | Area(km$^2$) | Population(people) |
| --- | --- | --- | --- |
| Unclassified(Spars） | 357 | 11061.13 | 2561345 |
| Mature residential area(C0) | 63 | 326.2918 | 572609 |
| Under developed area(C1) | 25 | 82.10567 | 162647 |
| Scenic area(C2) | 155 | 973.0869 | 1878303 |
| Commercial and entertainment area(C3) | 129 | 2068.882 | 1911066 |
| Area of public management, science, education and culture(C4) | 267 | 918.993 | 3551985 |
| New residential area(C5) | 122 | 974.2913 | 1271898 |

## *4.3 Examination of the results of identification*

In order to test the accuracy of the results of DZoF model, the map of different functional areas in Beijing obtained from the study was compared with the map of land-use status quo in the overall urban planning of Beijing (2004-2020)[5] and Google map of the area. Some results of comparison are shown as Table 4.

**Table 4** Comparison and analysis between Recognition results and the status quo of figure

| Contrast area | A famous scenic place in Beijing: Shidu Scenic Zone |
| --- | --- |
| Contrast diagram | 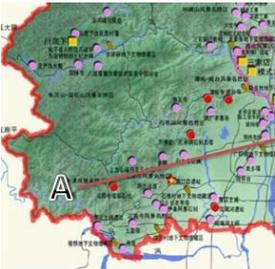 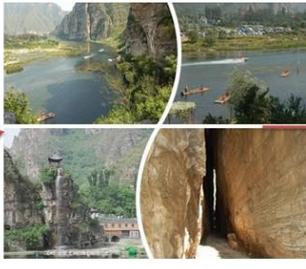 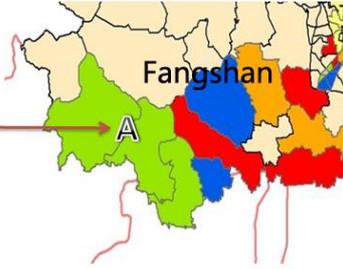 |
| Identification results | Region A(green) in the map of recognition results is a scenic area, which is in accord with region A in the map of land use status quo. |
| Contrast area | Culture relic protection site in Beijing: the ruins of the Turnoff City(which is 1.5km northwest of Badaling Guan City) |
| Contrast diagram | 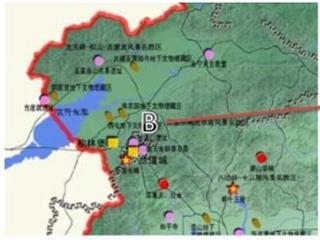 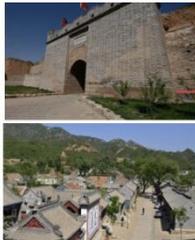 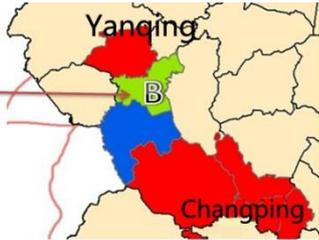 |

---

[5] The data is from the urban planning institute of Beijing

| | |
|---|---|
| Identification results | Region B(green) in the map of recognition results is a scenic area which corresponds to the location of the ruins of the Turnoff City |
| Contrast area | A delta in the watersheds of Yongding River in Beijing city which is mostly woodland. |
| Contrast diagram | 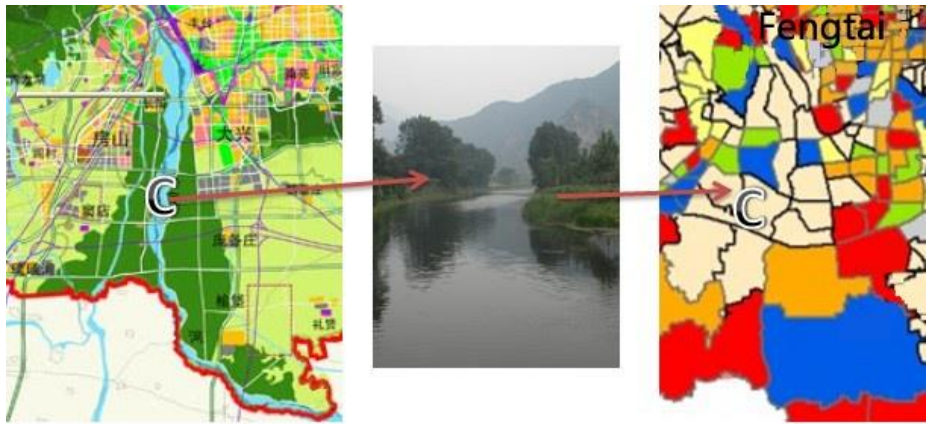 |
| Identification results | Region C(green) in the map of recognition results belongs to an unclassified area, according with region C in the map of land use status quo which represents the delta of Yongding River. |
| Contrast area | Haidian District has many universities including Peking University, Tsinghua University and Renmin University of China as well as Chinese silicon valley, which is the region of science, education and culture in Beijing. |
| Contrast diagram | 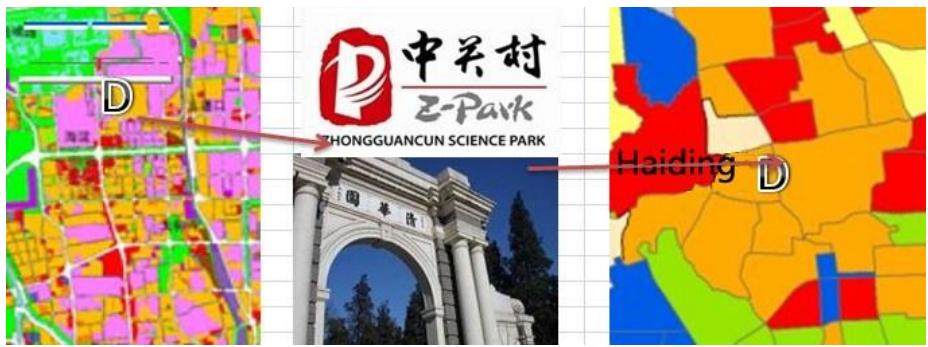 |
| Identification results | Region D(purple) in the map of land use status quo is the region of science, education and culture, which is in accord with region D in the map of recognition results. |
| Contrast area | some parts of Dongcheng District |
| Contrast diagram | 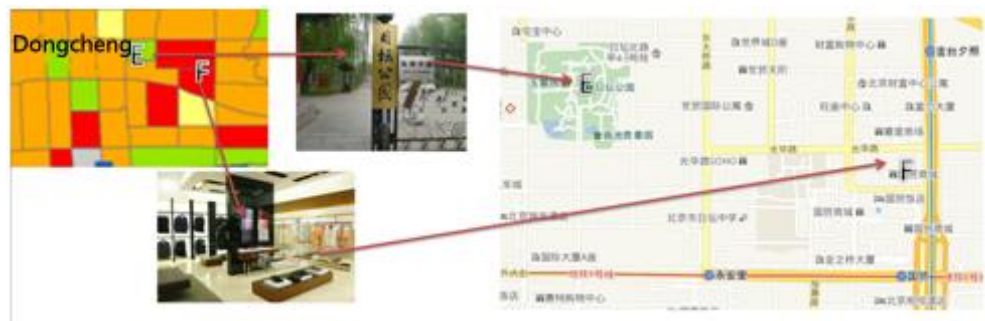 |
| Identification results | In the map of recognition results, region E is a scenic area and region F is a commercial and entertainment area. And by comparing with Google map of the regions, region E is Ritan Park in reality and region F is actually a famous commercial and entertainment area in Beijing, which has a plenty of shopping centers like China World Mall, Traders Hotel, Jiali Mall and Wanda Plaza. |

Besides, we also compared results of the study with detailed land-use data of each TAZ in Beijing to test the overall accuracy of identification. After ranking 1118 TAZ according to the size of public land (including land of public facilities and municipal administration), selecting the top 50 TAZ and removing the TAZ without SCD information, there were a total of 44 TAZ left. 22 of them were identified as areas of public management, science, education and culture with an accuracy rate of 50%. Using the same method, the residential land was analyzed and the accuracy rate was 58.06%. The result is as shown in table 5.

Table 5 Comparison and analysis between Recognition results and land use of TAZ

|  |  | Valid contrasting data | Identification results | Accuracy rate |
|---|---|---|---|---|
| Area of public management, science, education and culture | Number of TAZ | 44 | 22 | 50.00% |
|  | Total area($m^2$) | 59596348 | 19824488 |  |
| Residential area | Number of TAZ | 31 | 18 | 58.06% |
|  | Total area($m^2$) | 49319636 | 15423667 |  |

From the overall consideration of the highly-mixed land use status of Beijing and the contrastive analysis of the study, DZoF model has a certain degree of accuracy in effectively identifying main functional areas in Beijing.

## 5 Conclusion and discussion

This study is based on 77,976,010 records of SCD of Beijing City in one week in April 2008 and 113,810 pieces of POI data of Beijing in 2010. And by constructing a DZoF model, the study completed the identification of functional areas in Beijing and obtained 7 kinds of functional areas which is area of public management and culture, scenic area, commercial and entertainment area, mature residential area, new residential area and unclassified area. The area of public management and culture covers 267 transportation analysis zones (TAZ) with a total area of 918.993$km^2$. Business and entertainment area covers 129 TAZ with a total area of 2068.882$km^2$. And scenic area covers 155 TAZ with a total area of 973.0869 $km^2$.

The results show that DZoF model has certain ability to recognize the characteristic of functional areas in Beijing. This study can help people easily understand the spatial structure of a complex city, assist city planners to carry out planning of different urban functional areas based on human mobility and POI data, and provide the guidance and reference to city planning and site selection for real estate development.

This research has potential innovation in three aspects: first, it studied the spatial structure of a city through human mobility based on massive SCD; second, it combined the traditional methods of urban studies and big data mining and identified the features of residents' trip behavior from existing survey data of residents' trip, and applied these features as rules to the identification of urban functional area; third, it constructed a DZoF (discovering zones of different functions) model and identified the function of different regions in the city. As a whole, based on data of POI and SCD, the research studied the dynamic spatial structure of city using data mining method, providing a new analysis and research methods for studies of metropolitan spatial structure.

In this study, there are still some deficiencies that need to be improved in further researches: (1) the number of pick-up and drop-off passengers was chosen as an index in the study. But because of the unobtainable drop-off data of one-ticket lines, the study has ignored the data of one-ticket lines. New model can be built to effectively use data of one-ticket lines and improve the accuracy of recognition results in the future; (2) in 2010, the ratio of travelling by public bus in Beijing is 28.2%, while rail transit accounts for 11.5%, taxi for 6.6% and car for 34.2%[6]. So the data of bus have some limitations and one sidedness. In future researches, data of rail transit and taxi data can be added into the study to complete the information of human mobility and get more accurate results. (3) In real situation, mixed land use of commercial and residential services is widespread. The study ignored the possibility of mixed use and selected the function with the largest percentage to represent the region when collecting experimental results on the TAZ scale. Classification of mixed land use can be added into the future study.

---

[6] The data is from the 2011 report of transportation development of Beijing